\documentclass[a4paper,12pt]{article}
\usepackage{eurosym}
\usepackage[dvips]{graphicx}
\usepackage{amsfonts}
\usepackage{mathrsfs}
\usepackage{graphicx} 
\usepackage{epsfig}
\usepackage{amsmath, amsthm}
\usepackage{amssymb}

\newcommand{\ba}{\begin{array}}
\newcommand{\ea}{\end{array}}

\newcommand{\beq}{\begin{equation}}
\newcommand{\eeq}{\end{equation}}
\newcommand{\ben}{\begin{enumerate}}
\newcommand{\een}{\end{enumerate}}
\newcommand{\bit}{\begin{itemize}}
\newcommand{\eit}{\end{itemize}}

\begin{document}
\title{Integrable Flows of Curves/Surfaces, Generalized Heisenberg Ferromagnet  Equation and Complex Coupled Dispersionless  Equation}
\author{Guldana Bekova, Kuralay Yesmakhanova,  Gaukhar Shaikhova, \\ Gulgassyl Nugmanova  and Ratbay Myrzakulov\\
\textsl{Eurasian International Center for Theoretical Physics and} \\ { Department of General \& Theoretical Physics, Eurasian}  \\ National University,
Astana, 010008, Kazakhstan
}
\maketitle

\begin{abstract}
In the present paper, we study the Myrzakulov-XIII (M-XIII) equation   geometrically. From the
geometric point of view, we establish a link of the M-XIII equation  with the motion of space curves in the 3-dimensional space $R^{3}$. We also show that
 the complex coupled dispersionless   (CCD) equation  can be derived
 from
 the geometrical formalism  such that their curve flows are
 formulated. Finally, the gauge  equivalence between the M-XIII equation and the CCD equation is established.
\end{abstract}
\section{Introduction}
One of classical nonlinear differential  equations integrable by  through inverse scattering transform (IST) is the Heisenberg ferromagnet (HF) equation
\begin{equation}
iA_{t}+\frac{1}{2}[A, A_{xx}]=0.  \label{HF}
\end{equation}
It  describes the evolution of spin  waves in  nonlinear dispersive magnetic media. It admits several integrable and nonintegrable generalizations like the Landau-Lifshitz equation, Ishimori equation and so on. It  has been very successful applications  in 
physics  and mathematics [4]. 
However, in the different nonlinear regimes  of spin waves and for mathematical needs, the HF equation becomes less accurate or 
needs some extensions [6].  There are usually some approaches
to meet these requirements in the literature. For example, the first one is to add several
higher-order dispersive terms to get higher-order HF equation [2]. The second
one is to construct some multidimensional extensions. Several HF equations both integrable and nonintegrable  have been proposed by these approaches  [7, 8, 9, 10]. So that since the time the complete integrability of HF equations was discovered, many attempts for
constructing its generalization have been made [22]. 
One of such extensions of the HF equation is the M-XIII  equation which we are going to investigate in this paper.

The outline of the present paper is organized as follows. In section 2, we
present  the M-XIII equation. In section 3, the relation between the motion of space curves and the M-XIII equation is established. Then using this relation we found that the Lakshmanan (geometrical) equivalent counterpart of the M-XIII equation is the well-known complex coupled dispersionless (CCD)  equation. The gauge equivalence between the M-XIII equation and the CCD equation is established  in section 4. In Section 5, we study the relation between the M-XIII equation and differential geometry of surface.  The paper is concluded by comments and remarks in
section 6.

\section{Myrzakulov-XIII equation}

In this section, we consider the following Myrzakulov-XIII (M-XIII) equation
\begin{eqnarray}
iA_{s}=\frac{1}{2k}[A,A_{s}]_{y}+\frac{i}{k^{2}}(\rho A)_{y}.\label{5.15} 
\end{eqnarray}
Here $\sigma=\pm 1, \quad k=const$, 
\begin{eqnarray}
A=\left(
\begin{array}{cc}
A_{3} & \sigma A^{-} \\
A^{+} & -A_{3}
\end{array}
\right), \quad A^{2}=I, \quad A^{\pm}=A_{1}\pm i A_{2}, \quad \sigma(A_{1}^{2}+A^{2}_{2})+A^{2}_{3}=1
\end{eqnarray}
and
\begin{eqnarray}
  \rho=\pm\sqrt{1-\frac{\sigma k^{2}}{2}tr(A_{s}^{2})}=\sqrt{1-\sigma k^{2}{\bf A}_{s}^{2}}, \quad {\bf A}=(A_{1}, A_{2}, A_{3}), \quad {\bf A}^{2}=1.\label{5.15} 
\end{eqnarray}
The M-XIII equation (2) is integrable. Its Lax representation (LR) reads as
\begin{eqnarray}
\Phi_{y}&=&i(k-\lambda)A\Phi=U_{1}\Phi,\label{5.13} \\ 
\Phi_{s}&=&\left[\frac{i(k-\lambda)}{4k\lambda}\rho A+\frac{(k-\lambda)}{2\lambda}AA_{s}\right]\Phi=V_{1}\Phi.\label{5.14}
\end{eqnarray}

\section{Lakshmanan (geometrical) equivalent counterpart}
Let us now we find the Lakshmananan or that is same the geometrical equivalent counterpart of the M-XIII equation (2) for the case $\sigma=1$. To do that, let us rewrite the M-XIII equation (2) in the vector form. We have several equivalent vector forms:

i) 
\begin{eqnarray}
{\bf A}_{s}&=&\frac{1}{k}({\bf A}\wedge {\bf A}_{s})_{y}+\frac{1}{k^{2}}(\rho {\bf A})_{y},\\ \label{5.15} 
\rho_{y}&=&k {\bf A}\cdot ({\bf A}_{s}\wedge {\bf A}_{y}). \label{5.15} 
\end{eqnarray}

ii) 
\begin{eqnarray}
{\bf A}_{s}&=&\frac{1}{k}{\bf A}\wedge {\bf A}_{sy}+\frac{1}{k^{2}}\rho {\bf A}_{y},\\ \label{5.15} 
\rho_{y}&=&k {\bf A}\cdot ({\bf A}_{s}\wedge {\bf A}_{y}). \label{5.15} 
\end{eqnarray}

iii) 
\begin{eqnarray}
{\bf A}_{s}=\frac{1}{k}{\bf A}\wedge {\bf A}_{sy}+\frac{1}{k^{}}\partial_{y}^{-1}\left[{\bf A}\cdot ({\bf A}_{s}\wedge {\bf A}_{y})\right]{\bf A}. \label{5.15} 
\end{eqnarray}

iv)
\begin{eqnarray}
{\bf A}_{s}=\frac{1}{k}{\bf A}\wedge {\bf A}_{sy}\pm\frac{1}{k^{2}}\sqrt{1-\sigma k^{2}{\bf A}_{s}^{2}} {\bf A}_{y}. \label{5.15} 
\end{eqnarray}

Let us now we consider a  curve in  $R^{3}$ which is given by the unit   vectors ${\bf l}_{k}$. These vectors obey the Frenet-Serret  equations 
\begin{eqnarray}
\left ( \begin{array}{ccc}
{\bf  l}_{1} \\
{\bf  l}_{2} \\
{\bf  l}_{3}
\end{array} \right)_{y} = C
\left ( \begin{array}{ccc}
{\bf  l}_{1} \\
{\bf  l}_{2} \\
{\bf  l}_{3}
\end{array} \right),\quad
\left ( \begin{array}{ccc}
{\bf  l}_{1} \\
{\bf  l}_{2} \\
{\bf  l}_{3}
\end{array} \right)_{s} = G
\left ( \begin{array}{ccc}
{\bf  l}_{1} \\
{\bf  l}_{2} \\
{\bf  l}_{3}
\end{array} \right). \label{2.1} 
\end{eqnarray}
Here ${\bf l}_{1}, {\bf l}_{2}$ and ${\bf l}_{3}$ are the unit tangent, normal 
and binormal vectors  to the  curve respectively,  $x$ is its arclength 
parametrising the curve. The matrices $C$ and $G$ have the forms
\begin{eqnarray}
C =
\left ( \begin{array}{ccc}
0   & \kappa     & 0 \\
-\kappa & 0     & \tau  \\
0   & -\tau & 0
\end{array} \right) ,\quad
G =
\left ( \begin{array}{ccc}
0       & \gamma_{3}  & -\gamma_{2} \\
-\gamma_{3} & 0      & \gamma_{1} \\
\gamma_{2}  & -\gamma_{1} & 0
\end{array} \right).\label{2.2} 
\end{eqnarray}
The   curvature and torsion of the  curve  are given  by the following formulas
\begin{eqnarray}
\kappa= \sqrt{{\bf l}_{1y}^{2}},\quad \tau= \frac{{\bf l}_{1}\cdot ({\bf l}_{1y} \wedge {\bf l}_{1yy})}{{\bf l}_{1y}^{2}}.        \label{13}
\end{eqnarray}
The compatibility condition of the equations (13) is given by
\begin{eqnarray}
C_{s} - G_{y} + [C, G] = 0\label{14} 
\end{eqnarray}
or in elements   
 \begin{eqnarray}
\kappa_{s}    & = & \gamma_{3y} + \tau \gamma_2, \label{15} \\ 
\tau_{s}      & = & \gamma_{1y} - \kappa\gamma_2, \\ \label{16} 
\gamma_{2y} & = & \tau \gamma_3-\kappa \gamma_1. \label{17} 
\end{eqnarray}

Now we do the following identification:
 \begin{eqnarray}
{\bf A}\equiv {\bf l}_{1}. \label{18} 
\end{eqnarray}
Then we have 
\begin{eqnarray}
\kappa^{2} & = & {\bf A}_{y}^{2},\\ 
\tau&=&  \frac{{\bf A}\cdot ({\bf A}_{y} \wedge {\bf A}_{yy})}{{\bf A}_{y}^{2}} \label{2.3}
\end{eqnarray}
and
\begin{eqnarray}
\gamma_{1} & = & \frac{\kappa\rho\tau+k\kappa_{sy}}{\kappa k(k+\tau)},\\ 
\gamma_{2}&=& -\frac{\kappa_{s}}{k}, \\
\gamma_{3} & = &\frac{\kappa\rho}{k^{2}} -\frac{\kappa\rho\tau+k\kappa_{sy}}{k^{2}(k+\tau)}.       \label{2.3}
\end{eqnarray}

The  equations for $\kappa$ and $\tau$ reads as
 \begin{eqnarray}
\kappa_{s}&=&[\frac{\kappa\rho}{k^{2}} -\frac{\kappa\rho\tau+k\kappa_{sy}}{k^{2}(k+\tau)}]_{y}-\frac{\tau\kappa_{s}}{k}, \label{2.5} \\ 
\tau_{s}&=&[\frac{\kappa\rho\tau+k\kappa_{sy}}{\kappa k(k+\tau)}]_{y}+\frac{(\kappa^{2})_{s}}{2k}.  \label{2.7} 
\end{eqnarray}
Now we introduce a new function $v$ as
\begin{eqnarray}
v=k\int \frac{\gamma_{3}}{\kappa}ds.  
\end{eqnarray}
It is not difficult to verify that the functions $\kappa$ and $v$ are solutions of the following  equations
 \begin{eqnarray}
\kappa_{sy}&=&\kappa v_{y}v_{s}-0.5\partial_{y}^{-1}[(|q|^{2})_{s}]  \kappa, \label{2.5} \\ 
v_{sy}&=&-\frac{\kappa_{s}v_{y}}{\kappa_{y}v_{s}}.  \label{2.7} 
\end{eqnarray}
Next we introduce a new  complex function $q(s,y)$ as 
\begin{eqnarray}
q=\kappa e^{iv}.  
\end{eqnarray}
As result, the  function $q $  satisfies the following equation
\begin{eqnarray}
q_{sy}+0.5\partial_{y}^{-1}[(|q|^{2})_{s}] q=0.
\end{eqnarray}
Let us rewrite this equation as
\begin{eqnarray}
q_{sy}-\rho q&=&0, \\
\rho_{y}+0.5(|q|^{2})_{s}&=&0.
\end{eqnarray}
It is nothing but the focusing complex coupled dispersionless (CCD) system            \cite{feng}-\cite{1603.00781}. Thus we have proved the Lakshmanan (geometrical) equivalence between the M-XIII equation (2) and the CCD equation (33)-(34).  Finally, some comments in order. From (33)-(34) follows that \cite{feng}-\cite{1603.00781}
\begin{eqnarray}
\rho^{2}+\sigma |q_{s}|^{2}=const
\end{eqnarray}
or for simplicity
\begin{eqnarray}
\rho^{2}+\sigma |q_{s}|^{2}=1.
\end{eqnarray}
Thus
\begin{eqnarray}
\rho=\pm \sqrt{1-\sigma |q_{s}|^{2}}.
\end{eqnarray}
Finally instead of the set (33)-(34) we have
\begin{eqnarray}
q_{sy}\mp \sqrt{1-\sigma |q_{s}|^{2}}q=0.
\end{eqnarray}
On the face of it, the set of equations (33)-(34) contents two dependent variables $q$ and $\rho$. But as follows from (36), in fact we  have only one dependent variable namely $q(y,s)$.

\section{Gauge equivalent counterpart}
In the section 2, we have proved that the M-XIII equation (2) and the CCD equation (33)-(34) is the lakshmanan/geometrically equivalent each to other. Let us we now show that these equations are gauge equivalent  each to other. Consider the gauge transformation $\Phi=g^{-1}\Psi$, where $g=\Psi|_{\lambda=k}$. Then it is not difficult to show that the function $\Psi$ satisfies the following set of equations
\begin{eqnarray}
\Psi_{y}&=&U_{2}\Psi,\label{5.13} \\ 
\Phi_{s}&=&V_{2}\Phi,\label{5.14}
\end{eqnarray}
where
\begin{eqnarray}
U_{2} =  -i\lambda \sigma_{3}+0.5\left(
\begin{array}{cc}
0 & -q \\
\bar{q} & 0
\end{array}
\right), \quad V_{2} =  \frac{i}{4\lambda}[\rho \sigma_{3}+\left(
\begin{array}{cc}
0 & q_{s} \\
\bar{q}_{s} & 0
\end{array}
\right)].
\end{eqnarray}
The compatibility condition of the equations (39)-(40) is equivalent to the CCD equation (33)-(34). It means that between the M-XIII equation (2) and the CCD equation (33)-(34) takes place the gauge equivalence.

\section{Integrable  surfaces related with the M-XIII equation}

In this section, our aim is to extablish the link between  the M-XIII equation (2) and differential geometry of surface.

\subsection{Case 1: ${\bf A}\equiv{\bf r}_{y}$}

Consider the identification
 \begin{eqnarray}
 {\bf A}\equiv{\bf r}_{y},
 \end{eqnarray}
where   ${\bf r}(y,s)$ is the position vector of the curve embedded on the surface. In terms of ${\bf r}$,  the M-XIII equation (2) converted to the equations
\begin{eqnarray}
{\bf r}_{s}&=&\frac{1}{k}{\bf r}_{y}\wedge {\bf r}_{sy}+\frac{1}{k^{2}}\rho {\bf r}_{y},\\ \label{5.15} 
\rho_{y}&=&k {\bf A}\cdot ({\bf A}_{s}\wedge {\bf A}_{y}). \label{5.15} 
\end{eqnarray}
This set can be rewritten as
\begin{equation} 
\textbf{r}_{ys}+k \textbf{r}_y \times \textbf{r}_{s}=0.
\end{equation}
Note that for the case $k=-1$, the last equation  was obtained  in \cite{1603.00781} and studied in detail. As  integrable system, Eq.(43)-(44) admits the following LR
\begin{eqnarray}
F_{y}&=&0.5i(k-\lambda)r_{y}F=U_{3}F,\label{46} \\ 
F_{s}&=&\left[\frac{i(k-\lambda)}{2k\lambda}\rho r_{y}+\frac{(k-\lambda)}{2\lambda}r_{y}r_{ys}\right]F=V_{3}F.\label{47}
\end{eqnarray}
This LR gives
\begin{eqnarray} 
ir_{sy}&=&\frac{1}{4k}[r_{y},r_{sy}]_{y}+\frac{i}{4k^{2}}\rho  r_{yy}, \\
\rho_{y}&=&-ik tr( r_{y}\cdot [ r_{yy},  r_{sy}]).
\end{eqnarray}
It is just the matrix form of the equation (43)-(44). Also we note that
\begin{equation} 
\rho=k{\bf r}_y \cdot{\bf r}_{s}=k\cos\theta,
\end{equation}
so that $\theta$ represents the angle between the vectors $\textbf{r}_y$ and $\textbf{r}_s$.
In what follows for simplicity we assume that $k=1$. In the $\sigma=1$, from (36) follows that 
\begin{equation}
\rho=cos\theta, \quad q_s= \sin \theta e^{-\mathrm{i} \omega},
\end{equation}
where $\theta$ and $\omega$ are some real functions.
These formulas give us \cite{1603.00781}
\begin{equation}
q=(\theta _{y}-\mathrm{i}\omega _{y}\tan \theta) e^{-\mathrm{i}\omega}\,.
\end{equation}
Thus finally for LR (39)-(40) we obtain the expressions \cite{1603.00781}
\begin{eqnarray}
U_{2} &=& -i\lambda \sigma_{3}+\left(
\begin{array}{cc}
0 & -\frac 12(\theta _{y}-\mathrm{i}\omega _{y}\tan \theta) e^{-\mathrm{i}\omega}  \\
\frac 12(\theta _{y}+\mathrm{i}\omega _{y}\tan \theta) e^{\mathrm{i}\omega} & 0
\end{array}%
\right),  \\
V_{2}&=&\frac{\mathrm{i}}{4\lambda} \left(
\begin{array}{cc}
\cos \theta & \sin \theta e^{-\mathrm{i}\omega } \\
\sin \theta e^{\mathrm{i}\omega } & -\cos \theta%
\end{array}%
\right)\,.
\end{eqnarray}
Let us we return to the surface. Its  fundamental forms  read as \cite{1603.00781}
\begin{eqnarray}
  \text{I}&=&dy^2 + 2 \cos \theta dy ds + ds^2\,,
\label{Form1}\\
 \text{II}&=& (\tan \theta) \omega_y dy^2 + 2 \sin \theta dy ds +(\sin \theta) \omega_s ds^2  \,.
\label{Form2}
\end{eqnarray}
Now we are ready to write the  Gauss-Weingarten equations of the surface. We have \cite{1603.00781}
 \begin{eqnarray}
 \textbf{r}_{yy} &=& (\cot \theta) \theta_y \textbf{r}_{y} -(\csc \theta) \theta_y \textbf{r}_{s} -(\tan \theta) \omega_y \textbf{N},  \label{Gauss_eq1} \\
     \textbf{r}_{ys} &=& \sin \theta \textbf{N}  \,,   \label{Gauss_eq2}\\
  \textbf{r}_{ss} &=& -(\csc \theta) \theta_s \textbf{r}_{y}+(\cot \theta) \theta_s \textbf{r}_{s} +(\sin \theta) \omega_s \textbf{N},  \label{Gauss_eq3} \\
 \textbf{N}_{y} &=& (\cot \theta + \csc \theta \sec \theta \omega_y) \textbf{r}_{y}-(\csc \theta \omega_y +\csc \theta ) \textbf{r}_{s}  \,, \label{Weigarten_eq1} \\
   \textbf{N}_{s} &=&  -(\csc \theta - \cot \theta \omega_s) \textbf{r}_{y}+(\cot \theta + \csc \theta \omega_s ) \textbf{r}_{s} \,. \label{Weigarten_eq2}
 \end{eqnarray}
The compatibility conditions of these equations gives us  the following  Mainardi-Codazzi equation
\begin{equation}
(\omega_s \cos \theta)_y=\left( \frac{\omega_y}{\cos \theta}\right)_s\,.
\label{59}
\end{equation}
At the same time, the Gaussian curvature of the surface reads as 
\begin{equation}
K=-\frac{(\tan \theta) \omega_{y}\omega _{s}+\sin\theta}{\sin \theta}.
\label{63}
\end{equation}
 The important formula follows from the Liouville-Beltrami form  of the  \textit{Theorema egregium} and has the form \cite{1603.00781} 
\begin{equation}
\theta _{ys}-\sin \theta -(\tan \theta) \omega_{y}\omega _{s}=0\,.
\label{61}
\end{equation}

Let us write  the position vector on the surface in the component form as
\begin{equation}
{\bf r}=(r_{1},r_{2}, r_{3})
\end{equation}
or in the  matrix form 
\begin{equation} 
r=r_{1} e_1 + r_{2}e_2+r_{3} e_3.
\end{equation}
Now following \cite{1603.00781} we introduce  new  three matrices of the forms
\begin{equation}
 T = \Phi^{-1} e_3 \Phi, \quad  N = \Phi^{-1}e_2 \Phi, \quad  B = \Phi^{-1} e_1 \Phi,
\end{equation}
where 
\begin{equation}
 e_{j}=\frac{1}{2\mathrm{i}}\sigma_{j}.
\end{equation}
It follows from the following well-known formula
\begin{eqnarray}
r_y&=& \left. \Phi^{-1} U_\lambda \Phi\right|_{\lambda=1},\\
r_s&=& \left. \Phi^{-1} V_\lambda \Phi\right|_{\lambda=1}.
\end{eqnarray}
Thus we finally have 
\begin{eqnarray} 
r_y&=& \Phi^{-1} e_3 \Phi = T,\\
r_s&=&  (\cos \theta) T + (\sin \theta \cos \omega) N + (\sin \theta \sin \omega) B,
\end{eqnarray}
where   $y$ plays a role of arc length of the curve. These equations give us the following equation for the position vector ${\bf r}$ \cite{1603.00781}:
\begin{equation} \label{r_curve}
{\bf r}_s= \rho {\bf r}_y +  {\bf r}_y \wedge {\bf r}_{ys}.
\end{equation}
It coincide with the $r$-form of the M-XIII equation (45) after some transformation. Finally we note that 
\begin{equation} \label{r_curve}
{\bf r}_{s}^{2}= {\bf r}_{y}^{2}=1
\end{equation}
or in the matrix form
\begin{equation} \label{r_curve}
 r^{2}_s= r^{2}_{y}=I.
\end{equation}
\subsection{Case 2: ${\bf A}\equiv{\bf r}_{s}$}

Now let us consider the another type identification namely the following  one
 \begin{eqnarray}
 {\bf A}\equiv{\bf r}_{s},
 \end{eqnarray}
where   ${\bf r}(y,s)$ is the position vector of the curve embedded on the surface and $s$ is the arclength parameter of the curve. Then  the M-XIII equation (9)-(10) takes the form
\begin{eqnarray} 
\textbf{r}_{ss}&=&k^{-1} \textbf{r}_{s}\wedge  \textbf{r}_{yss}+k^{-2}\rho{\bf r}_{sy}, \\
\rho_{y}&=&k {\bf r}_{s}\cdot ({\bf r}_{ss}\wedge {\bf r}_{sy}).
\end{eqnarray}
This equation defines some integrable surface in $R^{3}$. Note that Eq.(77)-(78) is integrable with  the following LR
\begin{eqnarray}
F_{y}&=&0.5i(k-\lambda)r_{s}F=U_{4}F,\label{5.13} \\ 
F_{s}&=&\left[\frac{i(k-\lambda)}{2k\lambda}\rho r_{s}+\frac{(k-\lambda)}{2\lambda}r_{s}r_{ss}\right]F=V_{4}F.\label{5.14}
\end{eqnarray}
The compatibility  condition of the equations (79)-(80) gives
\begin{eqnarray} 
ir_{ss}&=&0.5k^{-1}[r_{s},r_{ss}]_{y}+ik^{-2}\rho  r_{sy}, \\
\rho_{y}&=&-0.5ik tr( r_{s}\cdot [ r_{ys},  r_{ss}]).
\end{eqnarray}
It is just the matrix form of the equation (77)-(78). 

\section{Soliton solutions of the M-XIII equation}

Let us now present the simple 1-soliton solution of the M-XIII equation (2). We use the corresponding  solution of the CCD equation (33)-(34). Let the seed solution of the last equation has the form
\begin{eqnarray} 
q=0, \quad \rho=1.
\end{eqnarray}
Then we have
\begin{eqnarray}
U_{2}=  -i\lambda y\sigma_{3}, \quad V_{2}=\frac{i}{4\lambda} s\sigma_{3}.
\end{eqnarray}
The corresponding  Sym-Tafel formula is given by \cite{1603.00781}
\begin{equation} 
r=\left.\Phi^{-1}\Phi_{\lambda}\right|_{\lambda=1}.
\end{equation}
Then the 1-soliton surface reads as \cite{1603.00781}
\begin{eqnarray}
  r_{1} &=& \frac{b}{(1-a)^2+b^2} \text{sech} R \cos W,  \\
  r_{2} &=& \frac{b}{(1-a)^2+b^2} \text{sech} R \sin W,  \\
  r_{3}&=& \frac{b}{(1-a)^2+b^2} \tanh R   +y + s,
\end{eqnarray}
where
\begin{equation*}
  R=b y + \frac{b}{a^2+b^2}s, \quad W=(1-a) y +\left(1+ \frac{a}{a^2+b^2} \right)s.
\end{equation*}
Finally we can write the 1-soliton solution of the M-XIII equation (2) as
\begin{equation} 
A_{1}=r_{1y}, \quad A_{2}=r_{2y}, \quad A_{3}=r_{3y}.
\end{equation}

\section{Conclusions}
 
In this paper, we have established the relation between the M-XIII equation (2) and the CCD equation (33)-(34). We have shown that the M-XIII equation (2) and the CCD equation (33)-(34) is the geometrically equivalent each to other. Also the gauge  equivalence between these equations is proved. 
Our results are significant for the deep understand
of integrable spin systems and their relations with differential geometry of curves and surfaces. 
 
 \section{Acknowledgements}
This work was supported  by  the Ministry of Edication  and Science of Kazakhstan under
grants 0118РК00935 and 0118РК00693.

 \end{document}